# Computational Microscopy beyond Perfect Lenses

Xingyuan Lu[1,2], Minh Pham[1,3], Elisa Negrini[3], Damek Davis[4], Stanley J. Osher[3], and Jianwei Miao[1*]

[1]*Department of Physics & Astronomy and California NanoSystems Institute, University of California, Los Angeles, CA 90095, USA.*

[2]*School of Physical Science and Technology, Soochow University, Suzhou 215006, China.*

[3]*Department of Mathematics, University of California, Los Angeles, CA 90095, USA.*

[4]*School of Operations Research and Information Engineering, Cornell University. Ithaca, NY 14850, USA.*

**We demonstrate that in situ coherent diffractive imaging (CDI), which harnesses the coherent interference between a strong and a weak beam illuminating a static and dynamic structure, can be a very dose-efficient imaging method. At low doses, in situ CDI can achieve higher resolution than perfect lenses with the point spread function as a delta function. Both our numerical simulation and experimental results show that the combination of in situ CDI and ptychography can reduce the dose by two orders of magnitude over ptychography. We expect that computational microscopy based on in situ CDI can be implemented in different imaging modalities with photons and electrons for low-dose imaging of radiation-sensitive materials and biological samples.**

The resolution of microscopy based on incoherent imaging is set by the point spread function (PSF) [1]. A microscope with the PSF as a delta function, which corresponds to an infinite numerical aperture without aberration, would represent a perfect imaging system. The other important factor for any practical microscope is the illumination, i.e., the number of particles (such as photons and electrons) per unit area and time required to form an image. With unlimited illumination, a phase-contrast microscope with perfect lenses would be an ultimate imaging system. However, many important systems such as energy materials, catalysts, polymers, and biological samples are very sensitive to x-rays or electrons, which is known as radiation damage effects [2-4]. In this Letter, we use mathematical analysis and numerical experiments to show that at low doses, computational microscopy based on in situ CDI can outperform microscopy with perfect lenses. CDI is a lensless imaging technique with the resolution only limited by the diffraction signal [5]. Over the years, CDI methods, such as conventional CDI, Bragg CDI and ptychography, have been broadly implemented using synchrotron radiation, x-ray free electron lasers, high harmonic generation, electron and optical microscopy [6-26]. However, although CDI methods have overcome the resolution limit set by the lens, their application to radiation-sensitive samples remains challenging. An approach to improve CDI's dose-efficiency is to incorporate strong scatterers in the vicinity of weakly scattering or biological samples [27-32]. More recently, numerical experiments have indicated that in situ CDI can significantly reduce the radiation dose by 1-2 orders of magnitude over conventional CDI [33]. However, the mathematical foundation on the radiation dose reduction of in situ CDI is unknown. Here we use mathematical analysis and numerical experiments to demonstrate that in situ CDI could be the most dose-efficient imaging method and at low doses, can achieve higher resolution than microscopy with perfect lenses.

In situ CDI takes advantage of the coherent interference from a static and a dynamic structure [Fig. 1(a)]. Let's define $\Psi_S$ and $\Psi_D$ as the Fourier transform of the static and the dynamic structure, respectively, where for simplicity we omit the reciprocal space coordinates and the time variable in the dynamic structure. We first consider conventional CDI without the static structure. The diffraction intensity of the dynamic structure is,

$$I_0 \left(\frac{r_e \lambda}{l_D \sigma_D}\right)^2 |\Psi_D|^2 = I_D \quad (1)$$

where $I_0$ is the fluence (photons per unit area) of the incident wave on the dynamic structure, $r_e$ is the classical electron radius, $\lambda$ is the wavelength, $I_D$ is the diffraction intensity (photons per pixel), $l_D$ and $\sigma_D$ are the size and the linear oversampling ratio of the dynamic structure [34], respectively. We rewrite Eq. (1) as,

$$\widetilde{D} \equiv \frac{r_e \lambda \sqrt{I_0}}{l_D \sigma_D} \Psi_D, \qquad \left|\widetilde{D}\right|^2 = I_D \quad (2)$$

Eq. (2) can be represented as a circle in the complex plane [Fig. 2(a)]. Although Eq. (2) is a non-convex problem, a unique solution [the red dot in Fig. 2(a)] can be solved by iterative phase retrieval algorithms [35], provided that the fluence is sufficiently high and the oversampling condition is satisfied [34]. However, when the fluence is very low, the circle becomes an annulus with the width of the ring set by Poisson noise in the diffraction intensity [Fig. 2(b)]. The presence of very high noise will make iterative algorithms trapped in local minima instead of converging to the global minimum [35].

An approach to overcome this major limitation is in situ CDI [Figs. 1(a) and Fig. S1 in Supplemental Material [36]], which is mathematically represented by,

$$\left|\frac{r_e \lambda \sqrt{I_1}}{l_S \sigma_S} \Psi_S + \frac{r_e \lambda \sqrt{I_0}}{l_D \sigma_D} \Psi_D\right|^2 = I \quad (3)$$

where $I_1$ is the fluence of the incident wave on the static structure with $I_1 \gg I_0$, $I$ is the diffraction intensity, $l_S$ and $\sigma_S$ are the size and the linear oversampling ratio of the static structure [34], respectively. We rewrite Eq. (3) as,

$$\tilde{S} \equiv \frac{r_e \lambda \sqrt{I_1}}{l_S \sigma_S} \Psi_S, \qquad |\tilde{S} + \tilde{D}|^2 = I \qquad (4)$$

As $|\tilde{S}| \gg |\tilde{D}|$, Eq. (4) can be represented by the intersection of a large circle and a small annulus in the complex plane [Fig. 2(c)]. Due to very low relative noise in $I$, iterative phase retrieval algorithms can be used to find the global solution of $\tilde{S} + \tilde{D}$ [i.e. the red dot in Fig. 2(c)] instead of trapping in local minima. To mathematically solve Eq. (4), we define,

$$\tilde{S} = FS = F_r S + i F_i S, \qquad \tilde{D} = FD = F_r D + i F_i D \qquad (5)$$

where $F$ is the Fourier transform, $S$ and $D$ are the static and dynamic structure, $F_r S$ and $F_i S$ are the real and imaginary part of $FS$, respectively. Substituting Eq. (5) into Eq. (4) and ignoring the very small $|\tilde{D}|^2$ term, we have

$$F_r S \, F_r D + F_i S \, F_i D \approx \frac{1}{2}(I - |\tilde{S}|^2) \qquad (6)$$

where $S$ can be accurately retrieved from the diffraction pattern due to the illumination of a high fluence on the static structure. Let's assume the array size of $D$ and $I$ to be $n \times n$ and $N \times N$, respectively. Since $D$ is complex and $I$ is real, with $N^2 > 2n^2$, there are more independent equations than the unknown variables [36]. To solve Eq. (6), we convert it to the matrix form,

$$AD = b \qquad (7)$$

$$A = diag(F_r S) F_r + diag(F_i S) F_i \qquad b = \frac{1}{2}(I - |\tilde{S}|^2)$$

where $A$ is a diagonal matrix with dimension $N^2 \times N^2$, $D$ has dimension $N^2 \times 1$ after padding with zeros, and $b$ has dimension $N^2 \times 1$. Eq. (7) can be solved by, $D = (A^T A)^{-1} A^T b$, where $A$ has independent columns, superscripts $T$ and $-1$ represent the transpose and inverse of a matrix. Although matrix $(A^T A)^{-1} A^T$ may have a large condition number, the solution can be stabilized by adding a damping term or $l_2$ regularizer. Alternatively, Eq. (7) can be solved by the least square method, $\min \frac{1}{2} \|AD - b\|^2$, where $D$ is constrained by a support. This becomes a convex problem and can be solved by optimization methods. Figure S2 in Supplemental Material [36] show specific examples of solving $D$ with Eqs. (8) and (9), which are comparable to that obtained by a traditional iterative phase retrieval algorithm.

To compare in situ CDI with a perfect imaging system, we simulated Zernike phase-contrast microscopy with perfect lenses [Fig. 1(b) and Supplemental Material [36]]. In our numerical experiment, we first simulated a dynamic biological vesicle of protein complexes in water, consisting of 20 frames with a thickness of 100 nm and a pixel size of 11.4 nm (Video S1 and Table S1 in Supplemental Material [36]). Figure S3(a)-(d) and Video S2(a) in Supplemental Material [36] show representative noisy phase-contrast images of the biological vesicle using perfect lenses with a fluence of 3.5e5, 3.5e6, 3.5e7, and 3.5e8 photons/µm², respectively. To examine the effect of the pixel size on the resolution, we created biological vesicles with a pixel size of 1, 5 and 10 nm and calculated their phase-contrast images with perfects lenses (Fig. S4 in Supplemental Material [36]), indicating that at these low doses, the resolution is limited by the dose instead of the pixel size.

We then used the time-varying biological vesicle as the dynamic structure and a 20-nm-thick Au pattern as the static structure (Fig. S1 in Supplemental Material [36]). With an x-ray energy of 530 eV, we calculated a time-sequence diffraction patterns from the complex exit waves of both the dynamic and static structures using Eq. (4), whereas the fluence on the static structure is 1.4e11 photons/μm² and the fluence on the dynamic structure varies from 3.5e5, 3.5e6, 3.5e7 to 3.5e8 photons/μm². Poisson noise was added to the diffraction intensity with the oversampling ratio of the static and dynamic structure to be 8 [34]. To reconstruct both the static and dynamic structure, we applied iterative phase retrieval algorithms to Eq. (4), which is more accurate than Eq. (8). We first reconstructed the static structure of the 20-nm-thick Au pattern from the diffraction patterns by combining the hybrid input-output algorithm [37] with shrinkwrap [38]. We then applied the static structure as a time-invariant constraint to retrieve the amplitudes and phases of the dynamic biological vesicle at different fluences using the generalized proximal smoothing algorithm [39]. We further improved the static structure during the phase retrieval of the time-sequence diffraction patterns [33], in which the illumination function of the incident wave can be incorporated if desired. Figure S4(e)-(h) and Video S2(b) in Supplemental Material [36] show the representative phase images of in situ CDI with a fluence of 3.5e5, 3.5e6, 3.5e7, and 3.5e8 photons/μm², respectively. Using the Fourier ring correlation (FRC), we quantitatively compared the phase images between in situ CDI and perfect lenses [Fig. S4(i) and Table S2 in Supplemental Material [36]], indicating that in situ CDI produces slightly higher-resolution phase images than perfect lenses.

To study the effects of the sample thickness on perfect lenses and in situ CDI, we increased the thickness of the dynamic biological vesicle to 300 nm. Figure 3(a)-(d) and Video S2(c) in Supplemental Material [36] show the noisy phase-contrast images of perfect lenses with a fluence of 3.5e5, 3.5e6, 3.5e7, 3.5e8 photons/μm², respectively. In comparison, we used the same phase-retrieval procedure of in situ CDI to reconstruct the amplitudes and phases of the dynamic biological vesicle, whereas the fluence on the dynamic structure varies from 3.5e5, 3.5e6, 3.5e7 to 3.5e8 photons/μm² and the fluence on the static structure is fixed at 1.4e11 photons/μm². Figure 3(e)-(h) and Video S2(d) in Supplemental Material [36] show the representative phase images of the biological vesicle. FRC comparisons indicate that in situ CDI produces higher-resolution phase images than the perfect lenses [Fig. 3(i) and Table S2 in Supplemental Material [36]], which is due to the fundamental difference of the imaging mechanism between the two methods. In situ CDI takes advantage of the coherent interference from all the pixels in the static and dynamic structure, from which a global solution of the complex wave is reconstructed. For a given fluence, the increase of the sample thickness improves the signal to noise ratio of the reconstructed images. This is in contrast to phase-contrast microscopy with perfect lenses, which forms images locally, that is, every pixel is independent of other pixels.

As most cellular structures are thicker than 500 nm, we increased the thickness of the dynamic biological vesicle to 1 μm. With such a thick sample, the weak phase approximation in Zernike phase-contrast microscopy does not hold any more, but CDI methods do not have such a limitation. To reconstruct the thick biological sample by in situ CDI with the lowest possible dose, we reduced the x-ray fluence to 1.75e3, 3.5e3, 6.2e3 and 7.7e3 photons/μm² (Fig. S5 in Supplemental Material [36]). In particular, a fluence of 1.75e3 photons/μm² corresponds to 0.23 photon/pixel and a dose of 137.5 Gy. No conventional CDI method can perform successful phase retrieval under such an extremely low dose condition. But in situ CDI successfully reconstructed the phase images of the biological vesicle in all these low dose cases. This numerical experiment further confirms our mathematical analysis, that is, the arc of the larger circle intersected by the smaller circle in Fig. 2(c) is almost a line segment. Therefore, finding the correct phase, i.e., the red dot in Fig. 2(c), becomes a linear problem, which can be solved by a linear solver as demonstrated in Fig. S2 in Supplemental Material [36].

Next, we performed a numerical simulation on the potential dose reduction by combining in situ CDI with ptychography, termed low-dose CDI (LoCDI). Figure S6 in Supplemental Material [36] shows the schematic of LoCDI, where a strong and a weak beam with $E = 530$ eV illuminate a static structure and a biological sample, respectively. The static structure consists of a 20-nm-thick Au pattern and the biological sample is HeLa cells with sharp features and a thickness of 300 nm. A raster scan of 10×10 positions was performed on the static structure and the biological sample by the strong and weak beam, respectively. At each scan position, a coherent interference pattern was collected from the static structure and the biological sample. All the numerical simulation parameters are shown in Table S3 in Supplemental Material [36]. From the 100 diffraction patterns, the static structure and the biological sample as well as the probe function were simultaneously reconstructed by the extended ptychographic iterative engine (ePIE) [40] [Fig. 4(e)-(h)]. In comparison, the phase images of the biological sample with perfect lenses under the same doses are shown in Fig. 4(a)-(d). Quantitative comparisons using the FRC curves indicate that LoCDI produces superior phase images than perfect lenses at these doses [Fig. 4(m) and Table S2 in Supplemental Material [36]]. We also used ptychography to reconstruct the biological samples with the strong beam removed [Fig. 4(i)-(l)]. FRC comparisons indicate that LoCDI can reduce the dose by two orders of magnitude over ptychography [Fig. 4(n) and Table S2 in Supplemental Material [36]].

To further validate our mathematical analysis and numerical simulation, we conducted a LoCDI experiment, where a laser beam was split into two beams and a lens was used to separate the two beam paths near the focal plane [Fig. 5(a)]. One beam was attenuated by a filter and the other was un attenuated. The two beams then illuminated two pinholes each with a diameter of 100 μm, creating a strong and a weak probe. The separation between two pinholes was set to be 1 mm to avoid any interference between the two probes on the sample. The sample, a USAF pattern [Fig. 5(b)], was positioned immediately after the pinholes and was raster scanned across the two probes. At each scan position, a diffraction pattern from the illumination of two probes was measured by an EMCCD, which has 1024×1024 pixels with a pixel size of 13 μm and was place 76.7 mm after the sample. For each data set, a total of 34×34 scan positions were conducted with an overlap ratio of 66%. By using different filters, we collected four data sets each with a total fluence as 2.2e7, 1.7e8, 3.7e8, and 3.8e9 photons/mm$^2$, while the fluence on the strong probe was kept at 2.2e10 photons/mm$^2$. Figure 5(c)-(f) show the reconstructed images at four different fluences by ePIE. In comparison, we acquired four ptychographic data sets by blocking the unattenuated beam, while keeping all the other parameters the same. Figure 5(g)-(j) show the reconstructions by ePIE with the same fluence on the sample. Quantitative comparison using the FSC curves indicates that LoCDI can reduce the dose by two orders of magnitude over ptychography [Fig. 5(k) and Supplemental Table S2].

By harnessing the coherent interference between a strong and a weak beam illuminating a static and a dynamic structure, in situ CDI takes advantage of the benefits of both holography [41,42] and CDI [20,33] and is a very dose-efficient imaging method. Our numerical simulation and experimental results show that in situ CDI can not only achieve higher resolution than perfect lenses at low doses, but also be combined with ptychography to reduce the dose by two orders of magnitude over ptychography. As in situ CDI uses the static structure as the time-invariant constraint to reconstruct the dynamic structure which is time-variant, it can achieve high temporal resolution that is only limited by the fluence and detector read-out speed. In combination with coherent x-ray sources [20] and bright electron sources [43], we expect that in situ CDI will be a powerful dose-efficient imaging method to probe the structure and dynamics of a wide range of radiation-sensitive materials and biological samples.

This work was supported by the U.S. Air Force Office Multidisciplinary University Research Initiative (MURI) program under award no. FA9550-23-1-0281 and STROBE: a National Science Foundation Science and Technology Center under award no. DMR-1548924. M.P. and E.N. acknowledge the support by Simons Postdoctoral program at IPAM and DMS1925919. D.D. acknowledges the support by an Alfred P. Sloan research fellowship and NSF DMS award 2047637.

X.L. and M.P. contributed equally to this work. J.M. directed the project; M.P., J.M., E.N., D.D., S.J.O. discussed and/or carried out the mathematical analysis; X.L., M.P., E.N. and J.M. discussed and/or performed the numerical simulation; J.M. wrote the manuscript with input from X.L., M.P. and E.N. All authors commented on the manuscript.

*Email: miao@physics.ucla.edu

**Reference:**

1. E. Wolf (ed.), *Progress in Optics*, vol. 51, pp. 355, Elsevier (2008).

2. R. Henderson, Q. Rev. Biophys. **28**, 171 (1995).

3. M.. R. Howells et al., J. Electron Spectros. Relat. Phenomena **170**, 4 (2009).

4. M. Du, and C. Jacobsen, Ultramicroscopy **184**, 293 (2018).

5. J. Miao, P. Charalambous, J. Kirz, and D. Sayre, Nature **400**, 342 (1999).

6. J. Miao et al., Phys. Rev. Lett. **89**, 088303 (2002).

7. D. Shapiro et al., Proc. Natl. Acad. Sci. U.S.A. **102**, 15343 (2005).

8. M. A. Pfeifer et al., Nature **442**, 63 (2006).

9. H. N. Chapman *et al.*, Nat. Phys. **2**, 839 (2006).

10. J. M. Rodenburg et al., Phys. Rev. Lett. **98**, 034801 (2007).

11. P. Thibault et al., Science **321**, 379 (2008).

12. C. Song et al., Phys. Rev. Lett. **101**, 158101 (2008).

13. I. Robinson and R. Harder, Nat. Mater. **8**, 291 (2009).

14. K. Giewekemeyer et al. Proc. Natl. Acad. Sci. U.S.A. **107**, 529 (2009).

15. Y. Nishino et al., Phys. Rev. Lett. **102**, 018101 (2009).

16. H. N. Chapman, and K. A. Nugent, Nat. Photonics **4**, 833 (2010).

17. H. Jiang et al., Proc. Natl. Acad. Sci. U.S.A. **107**, 11234 (2010).

18. M. M. Seibert *et al.*, Nature **470**, 78 (2011).

19. G. Zheng, R. Horstmeyer, and C. Yang, Nat. Photonics **7**, 739 (2013).

20. J. Miao, T. Ishikawa, I. K. Robinson, and M. M. Murnane, Science **348**, 530 (2015).

21. D. F. Gardner *et al.*, Nat. Photonics **11**, 259 (2017).

22. M. Holler et al., Nature **543**, 402 (2017).

23. S. Gao et al., Nat. Commun. **8**, 163 (2017).

24. Y. Jiang et al., Nature **559**, 343 (2018).

25. F. Pfeiffer, Nat. Photon **12**, 9 (2018).

26. A. Rana et al., Nat. Nanotechnol. **18**, 227 (2023).

27. S. Boutet at al., J. Electron Spectros. Relat. Phenomena **166–167**, 65 (2008).

28. C. T. Putkunz et al., Phys. Rev. Lett. **106**, 013903 (2011).

29. T. Y. Lan, P. N. Li, and T. K. Lee, New J. Phys. **16**, 033016 (2014).

30. C. Kim et al., Opt. Express **22**, 29161 (2014).

31. Y. Takayama et al., Sci. Rep. **5**, 8074 (2015).

32. D. Noh et al., J. Phys. Condens. Matter **28**, 493001 (2016).

33. Y. H. Lo et al., Nat. Commun. **9**, 1826 (2018).

34. J. Miao, D. Sayre, and H. N. Chapman, J. Opt. Soc. Am. A **15**, 1662 (1998).

35. Y. Shechtman et al., IEEE Signal Processing Mag. **32**, 87 (2015).

36. See Supplemental Material at http://link.aps.org/ supplemental/XXX for supplemental figures, table and videos.

37. J. R. Fienup et al., Appl. Opt. **21**, 2758 (1982).

38. S. Marchesini et al., Phys. Rev. B **68**, 140101 (2003).

39. M. Pham, P. Yin, A. Rana, S. Osher, and J. Miao, Opt. Express **27**, 2792 (2019).

40. A. Maiden, D. Johnson, and P. Li, Optica **4**, 736–745 (2017).

41. T. Latychevskaia, J. Opt. Soc. Am. A **36**, D31-D40 (2019).

42. L. M. Lohse et al., J. Synchrotron Radiat. **27**, 852-859 (2020).

43. P. Musumeci et al., Nucl. Instrum. Methods Phys. Res. A **907**, 209 (2018).

**Figures and Figure Captions**

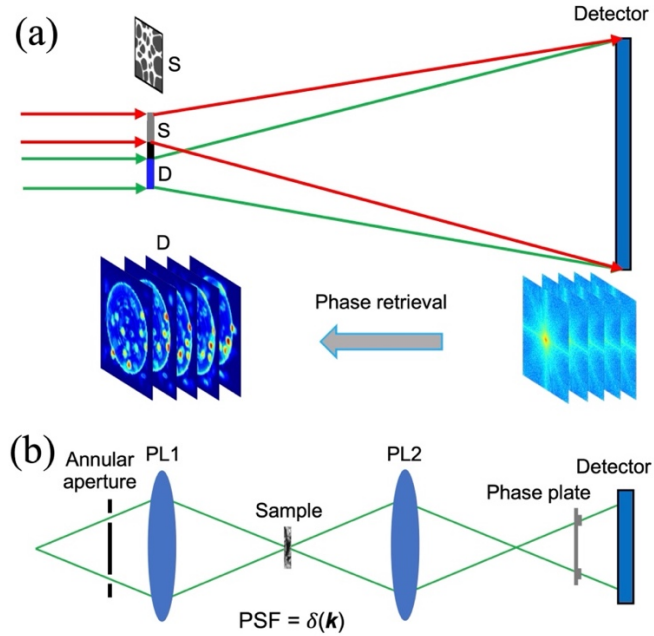

**Fig. 1.** Schematic of in situ CDI and phase-contrast microscopy with perfect lenses. (a) In situ CDI consists of a static (S) and a dynamic (D) structure that are coherently illuminated by a strong and a weak beam, respectively. Time-sequence diffraction patterns are collected from the coherent interference of the two beams by a detector, from which time-sequence phase images are reliably reconstructed by phase retrieval algorithms. (b) A perfect phase-contrast microscope with the PSF as a delta function, which consists of a source, an annular aperture, a first perfect lens (PL1), a sample, a second perfect lens (PL2), a phase plate, and a detector.

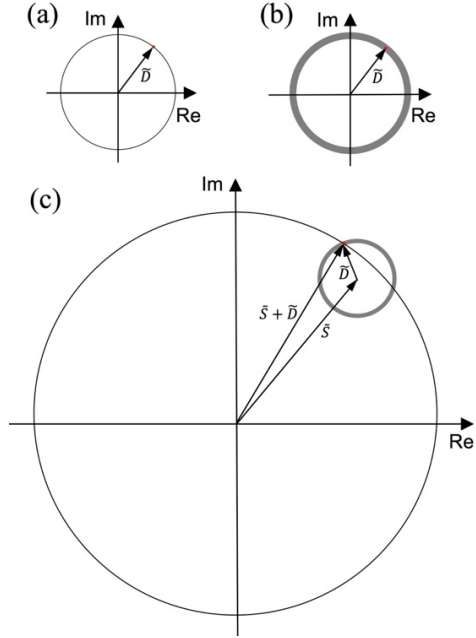

**Fig. 2.** CDI-based computational microscopy in the complex plane. (a) In conventional CDI, the Fourier component of each pixel in reciprocal space represents a vector, $\tilde{D}$, confined in a circle. With low noise and the oversampling condition satisfied [34], the correct phase (red dot) can be recovered by phase retrieval algorithms. (b) With very high noise, the circle becomes an annulus and phase retrieval algorithms are usually trapped in local minima instead of converging to the global minimum (red dot) [35]. (c) In in situ CDI, the coherent interference between a strong beam illuminating a static structure and a weak beam illuminating a dynamic structure produces a large circle in the complex plane, which intersects with a small annulus from the dynamic structure. Phase retrieval algorithms can quickly find the global minimum (red dot) instead of trapping in local minima as it becomes a convex problem, i.e., the intersected arc is almost a line segment.

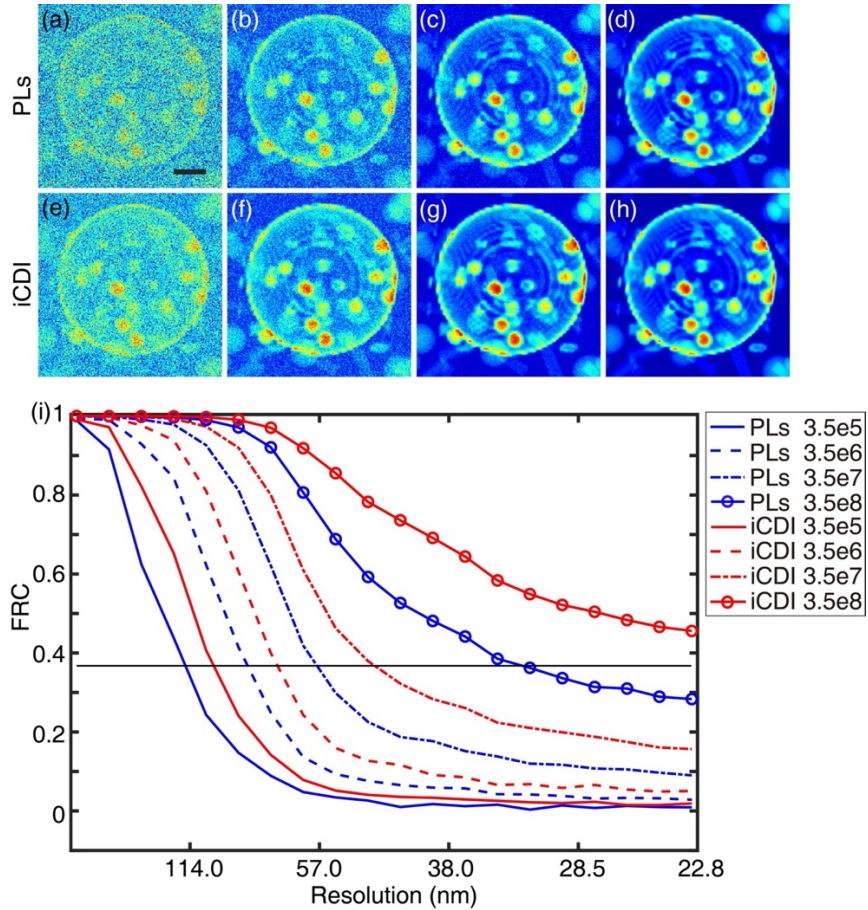

**Fig. 3.** Numerical experiments on phase-contrast microscopy with perfect lenses (PLs) and in situ CDI (iCDI). (a)-(d) Representative images of a 300-nm-thick dynamic biological vesicle obtained by phase-contrast microscopy with perfect lenses using an x-ray fluence of 3.5e5, 3.5e6, 3.5e7, and 3.5e8 photons/µm$^2$, respectively, corresponding to a dose of 2.75e4, 2.75e5, 2.75e6, and 2.75e7 Gy, respectively. Scale bar, 500 nm. (e)-(h) The same images reconstructed by in situ CDI with an x-ray fluence of 3.5e5, 3.5e6, 3.5e7, and 3.5e8 photons/µm$^2$, respectively, and a fixed fluence of 1.4e11 photons/µm$^2$ on the static structure. (i) Average Fourier ring correlation (FRC) curves of the 20-frame phase images obtained by perfect lenses and in situ CDI as a function of the x-ray fluence. The black line with FRC = 1/e indicates the resolution.

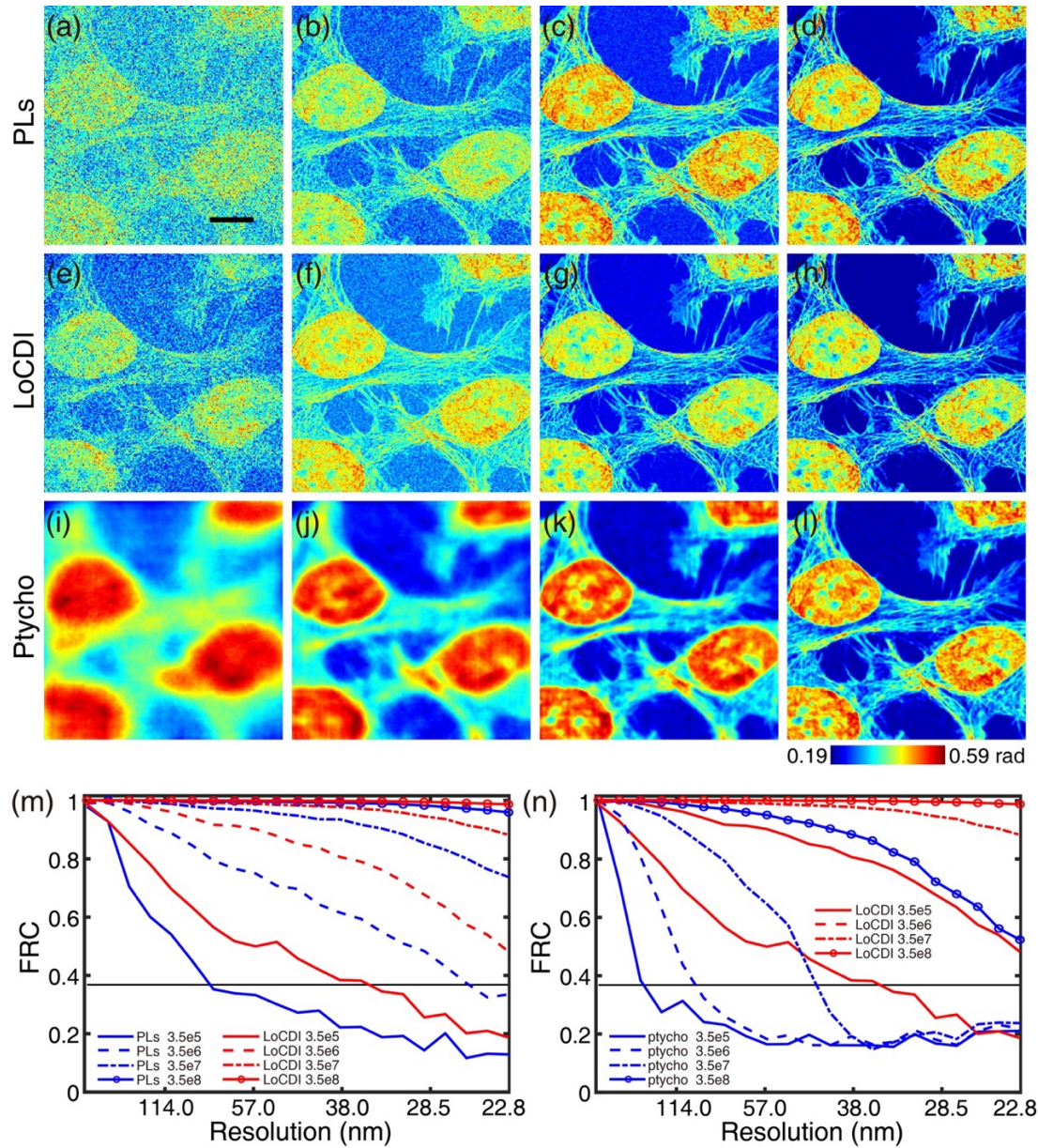

**Fig. 4**. Numerical simulation on the dose reduction with LoCDI. (a)-(d) Phase-contrast images with perfect lenses using an x-ray fluence of 3.5e5, 3.5e6, 3.5e7 and 3.5e8 photons/µm$^2$, respectively. Scale bar, 500 nm. (e)-(h) Phase images of LoCDI with an x-ray fluence of 3.5e5, 3.5e6, 3.5e7 and 3.5e8 photons/µm$^2$ on the biological sample, respectively, and a fixed fluence of 1.4e11 photons/µm$^2$ on the static structure. (i)-(l) Phase images of ptychography with the same x-

ray fluences as those of LoCDI, but without the strong beam. (m) FRC comparisons indicates that LoCDI is superior to phase microscopy with perfect lenses. (n) FRC comparisons with the ground truth of the biological sample indicate that LoCDI can reduce the dose by two orders of magnitude over ptychography with a fluence of 3.5e5 photons/μm$^2$.

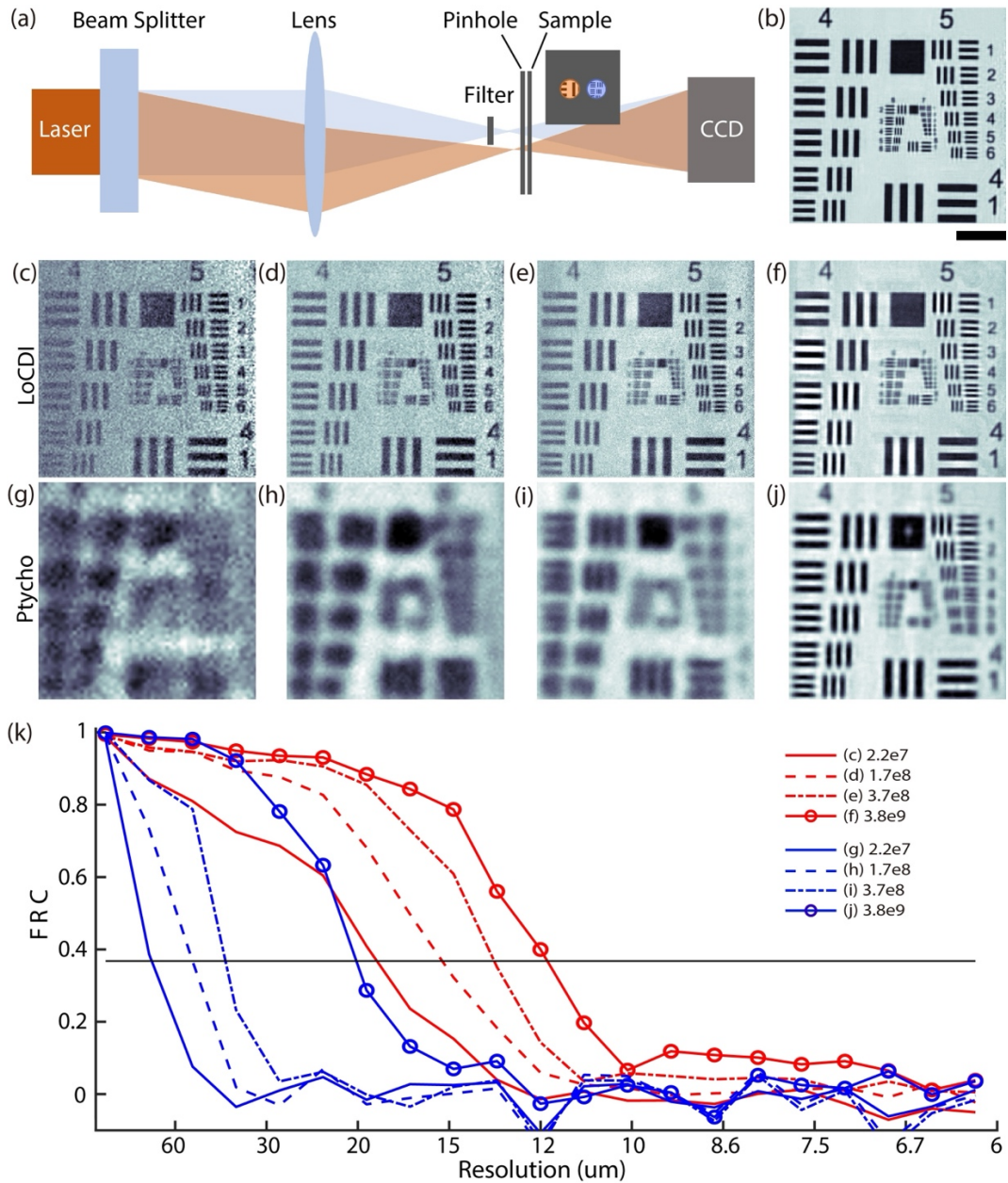

**Fig. 5**. Experiment demonstration on significant dose reduction with LoCDI. (a) Schematic of the experimental setup using a laser with a wavelength of 532 nm (CCD: charge-coupled device). (b) The structure of a USAF pattern, obtained by a high-dose ptychography experiment. Scale bar, 200 μm. (c)-(f) Reconstructed images with a fluence of 2.2e7, 1.7e8, 3.7e8, and 3.8e9 photons/mm$^2$, respectively, illuminating the sample (light blue), while the fluence of the unattenuated probe (brown) was kept at 2.2e10 photons/mm$^2$. (g)-(j) Ptychographic reconstructions of the same sample with a fluence of 2.2e7, 1.7e8, 3.7e8, and 3.8e9 photons/mm$^2$, respectively, while the unattenuated beam (brown) was blocked. (k) FRC curves of LoCDI and ptychographic reconstructions as a function of the fluence, which were calculated with the structure in (b).

# Computational Microscopy beyond Perfect Lenses

Xingyuan Lu[1,2], Minh Pham[1,3], Elisa Negrini[3], Damek Davis[4], Stanley J. Osher[3], and Jianwei Miao[1]

[1]*Department of Physics & Astronomy and California NanoSystems Institute, University of California, Los Angeles, CA 90095, USA.*
[2]*School of Physical Science and Technology, Soochow University, Suzhou 215006, China.*
[3]*Department of Mathematics, University of California, Los Angeles, CA 90095, USA.*
[4]*School of Operations Research and Information Engineering, Cornell University. Ithaca, NY 14850, USA.*

## Supplemental Text

**Phase-contrast microscopy with perfect lenses**

In Zernike phase-contrast microscopy, the phase-contrast intensity ($I_p$) of a weak phase object can be approximated as [44-46],

$$I_p = I_i(a^2 \pm 2a\varphi) \qquad (10)$$

where $a$ and $\varphi$ are the transmissivity and the phase shift of the sample (defined below), respectively, $I_i$ is the intensity of the incident beam, and the $\pm$ sign indicates a positive ($\frac{\pi}{2}$) and a negative ($-\frac{\pi}{2}$) phase shift.

$$a = exp(-\mu z/2) \quad \text{and} \quad \varphi = \frac{2\pi}{\lambda}\delta z \qquad (11)$$

where $\mu = \frac{4\pi}{\lambda}\beta$, $\beta = \frac{r_e\lambda^2}{2\pi}n_a f_2$, $\delta = \frac{r_e\lambda^2}{2\pi}n_a f_1$, $f_1$ and $f_2$ are the real and imaginary part of the atomic scattering factor, and $z$ is the thickness of the sample [46]. To calculate the phase-contrast images with perfect lenses, we convolved Eq. (10) with the PSF of perfect lenses, i.e., a delta function. Poisson noise was then added to each image based on the x-ray fluence.

44. F. Zernike, Science **121**, 345 (1955).
45. M. Born and E. Wolf, *Principles of Optics*, Pergamon Press, Oxford, 6[th] ed. (1980).
46. J. Kirz, C. Jacobsen, and M. Howells, Q. Rev. Biophys. **28**, 33-130 (1995).

**Supplemental Figures**

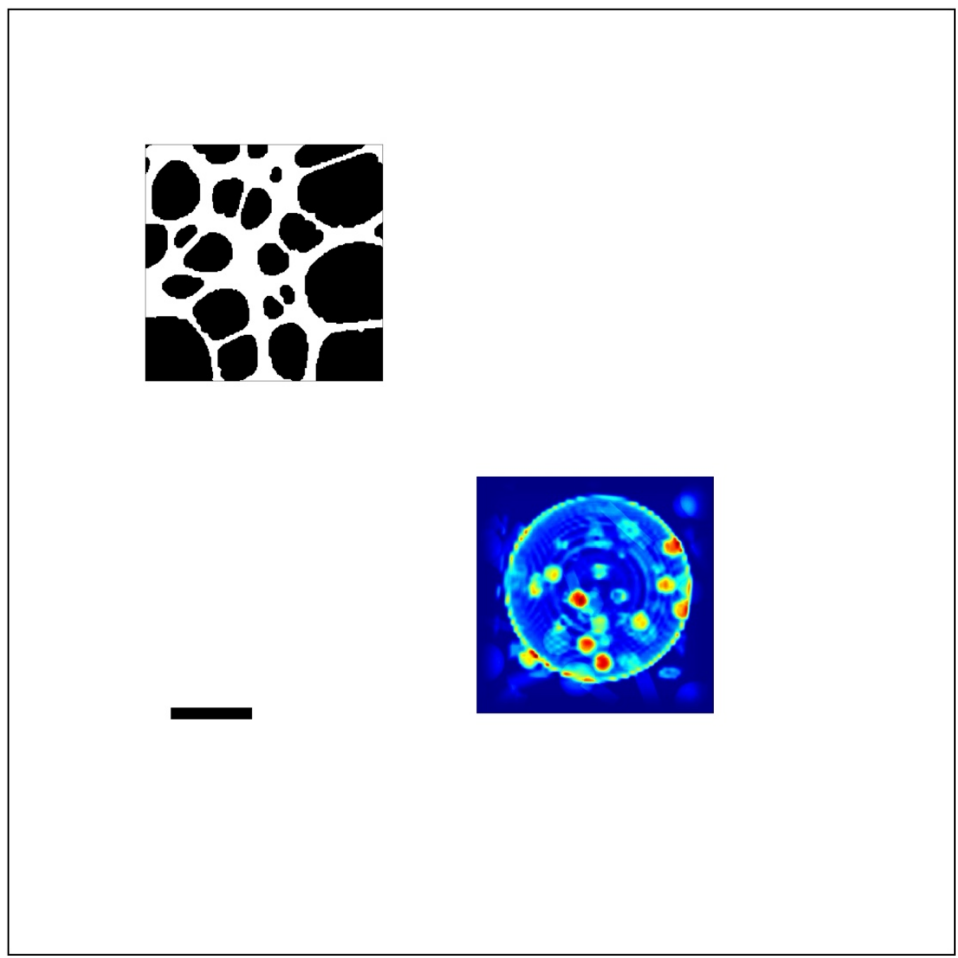

**Fig. S1.** Schematic of a static and a dynamic structure of in situ CDI used for the numerical experiments. The static structure consists of a 20-nm-thick Au pattern with a size of 2.92×2.92 µm$^2$. The dynamic structure is a 20-frame biological vesicle with a size of 2.92×2.92 µm$^2$ (Video S1). The black square indicates the whole reconstructed region. Scale bar, 1 µm.

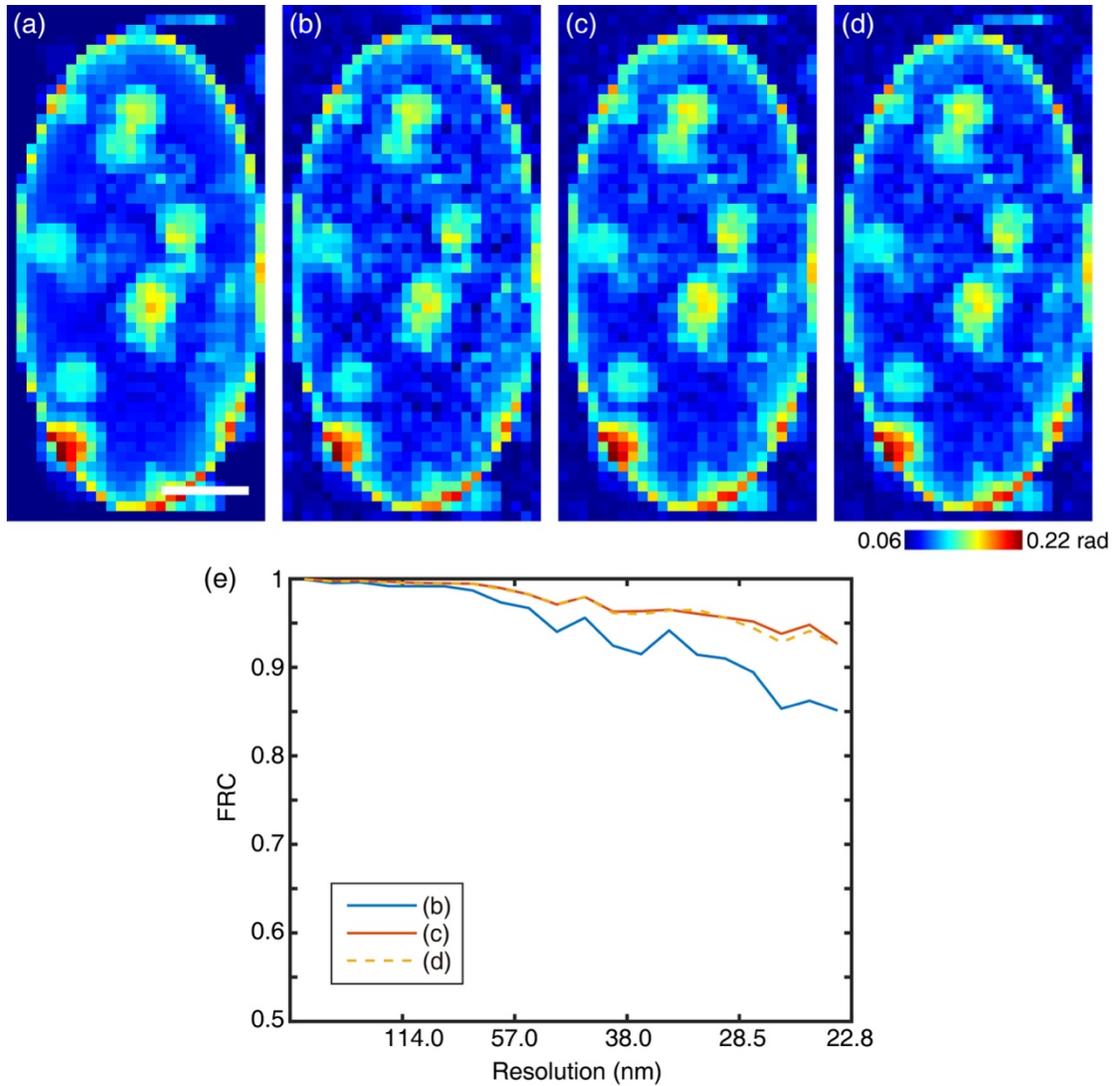

**Fig. S2.** (a) Model of a biological vesicle. (b) and (c) Reconstructed images by a linear system solver with Eq. (8) (b) and linearized alternating projection with Eq. (9) and $l_2$ regularization (c), which are comparable to that obtained by the generalized proximal smoothing algorithm [39] (d). The x-ray fluence on static and dynamic structure is 1.86e10 and 6.15e4 photons/µm$^2$, respectively. Scale bar, 100 nm.

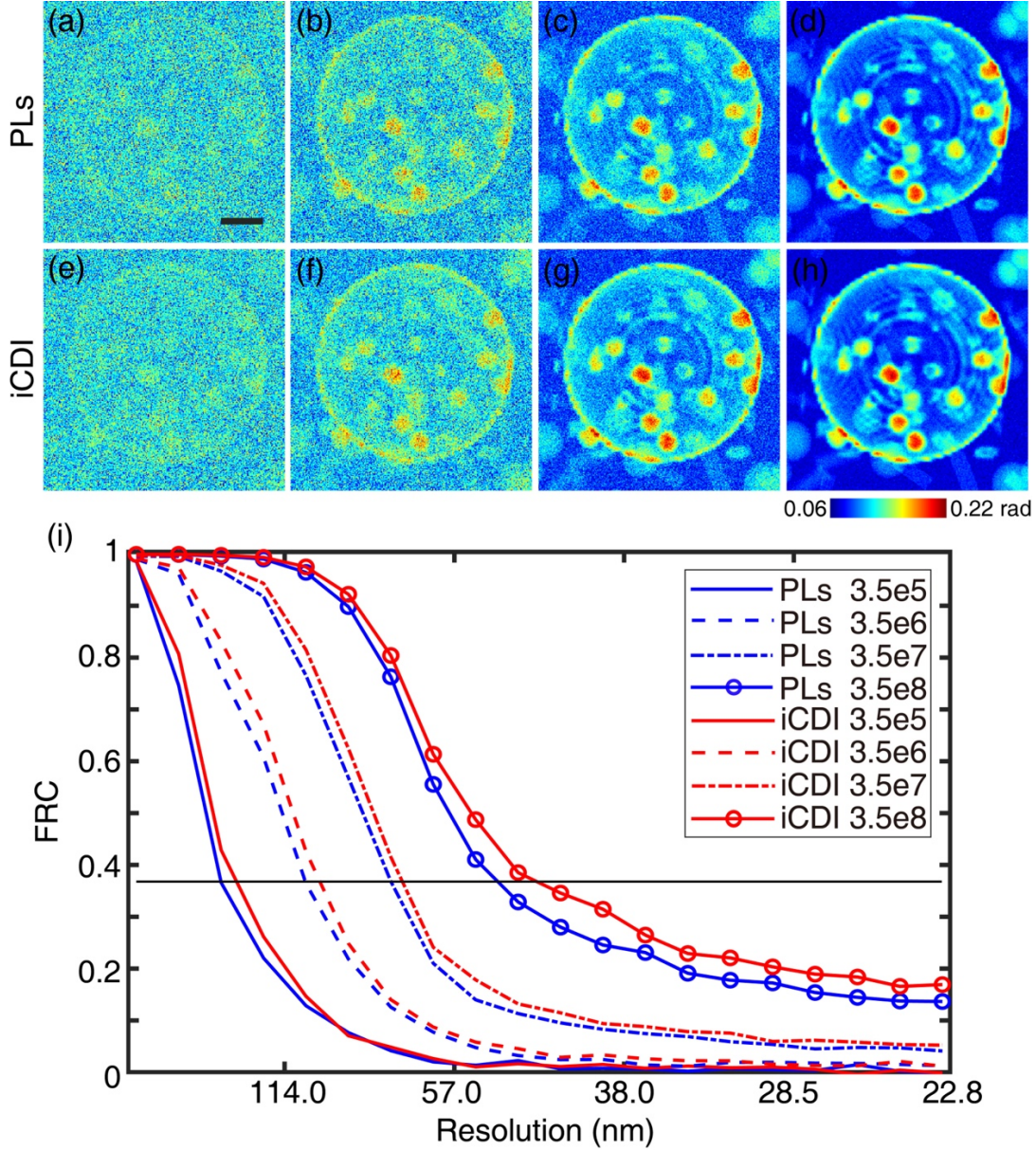

**Fig. S3.** Numerical experiments on phase-contrast microscopy with perfect lenses (PLs) and in situ CDI (iCDI). (a)-(d) Representative images of a 100-nm-thick dynamic biological vesicle obtained by phase-contrast microscopy with perfect lenses using an x-ray fluence of 3.5e5, 3.5e6, 3.5e7, and 3.5e8 photons/μm$^2$, respectively, corresponding to a dose of 2.75e4, 2.75e5, 2.75e6, and 2.75e7 Gy, respectively. (e)-(h) The same images reconstructed by in situ CDI with an x-ray fluence of 3.5e5, 3.5e6, 3.5e7, and 3.5e8 photons/μm$^2$, respectively, on the biological vesicle and a fixed fluence of 1.4e11 photons/μm$^2$ on the static structure. (i) Average Fourier ring correlation (FRC) curves of the 20-frame phase images obtained by perfect lenses and in situ CDI as a function of the x-ray fluence, where the FRC curves are calculated in comparison with the dynamic biological vesicle model. The black line with FRC = 1/e indicates the resolution (Table S2). Scale bar, 500 nm.

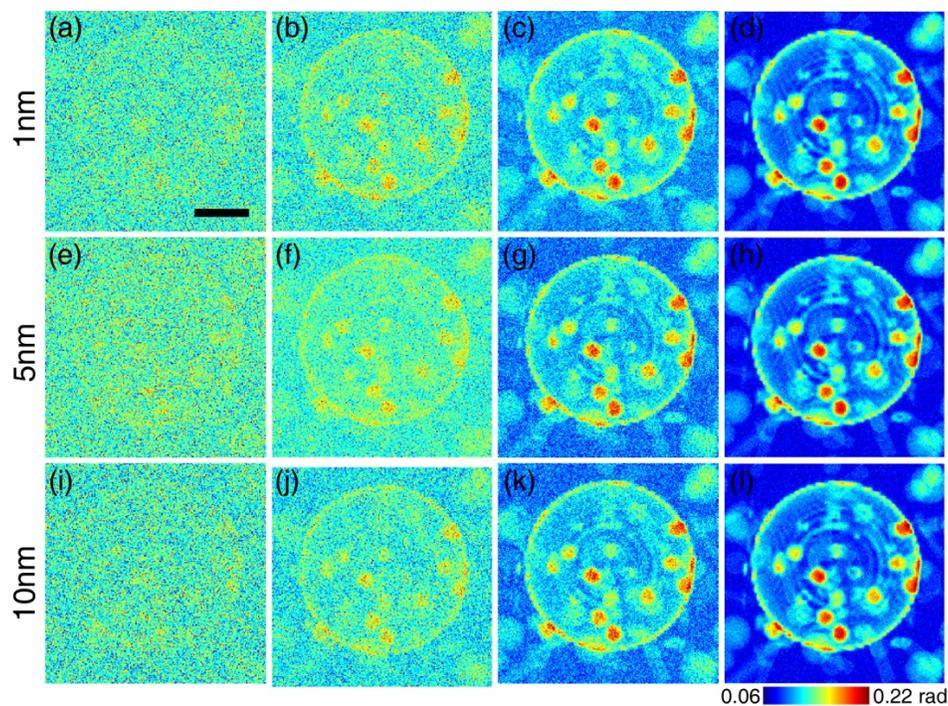

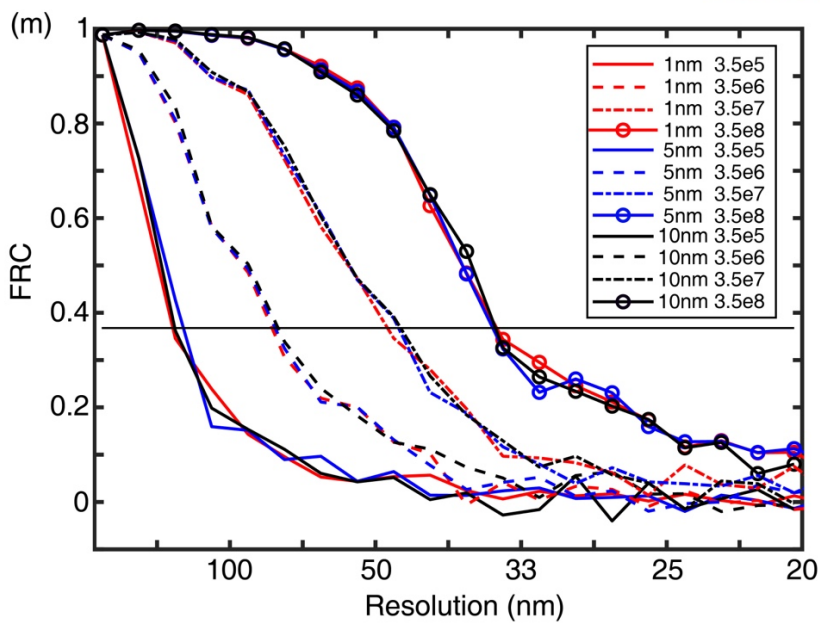

**Fig. S4**. The effect of the pixel size on the resolution. Zernike phase-contrast images of a biological vesicle with perfects lenses and a pixel size of 1 nm [(a)-(d)], 5 nm [(e)-(h)] and 10 nm [(i)-(l)]. The x-ray fluence on the biological vesicle is 3.5e5 [(a), (e), (i)], 3.5e6 [(b), (f), (j)], 3.5e7 [(c), (g), (k)], and 3.5e8 photons/$\mu m^2$ [(d), (h), (l)]. Some fine structural features were introduced in the vesicle model at these length scales. FRC comparisons between the phase-contrast images and the biological vesicle model, indicating that at these low doses, the resolution is limited by the dose instead of the pixel size. Scale bar, 500 nm.

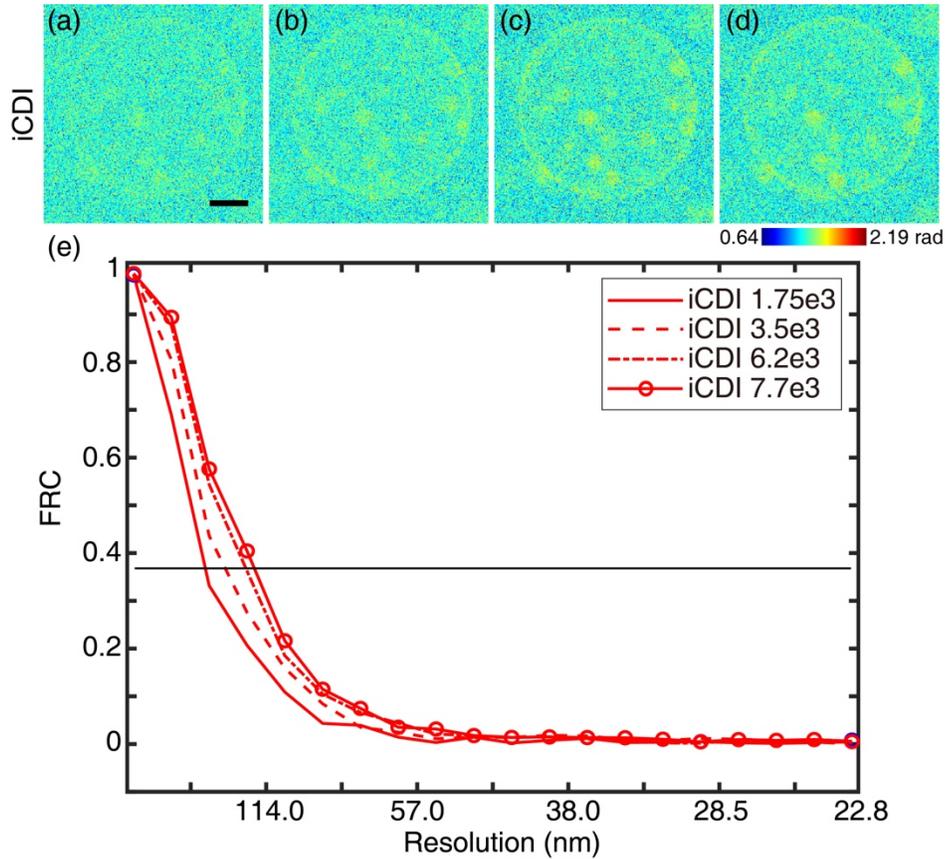

**Fig. S5.** Numerical experiments on the in situ CDI reconstruction of a thick biological sample with the lowest possible dose. (a)-(d) Representative images of a 1-µm-thick dynamic biological vesicle obtained by in situ CDI with an x-ray fluence of 1.75e3, 3.5e3, 6.2e3 and 7.7e3 photons/µm$^2$, respectively, on the biological vesicle and a fixed fluence of 1.4e11 photons/µm$^2$ on the static structure. Even with a fluence of 1.75e3 photons/µm$^2$ (corresponding to 0.23 photon/pixel), the biological vesicle is still visible in the phase image of in situ CDI. (e) Average FRC curves of the 20-frame phase images reconstructed by in situ CDI as a function of the fluence, where the FSC curves are calculated in comparison with the dynamic biological vesicle model. The black line with FRC = 1/e indicates the resolution (Table S2). Scale bar, 500 nm.

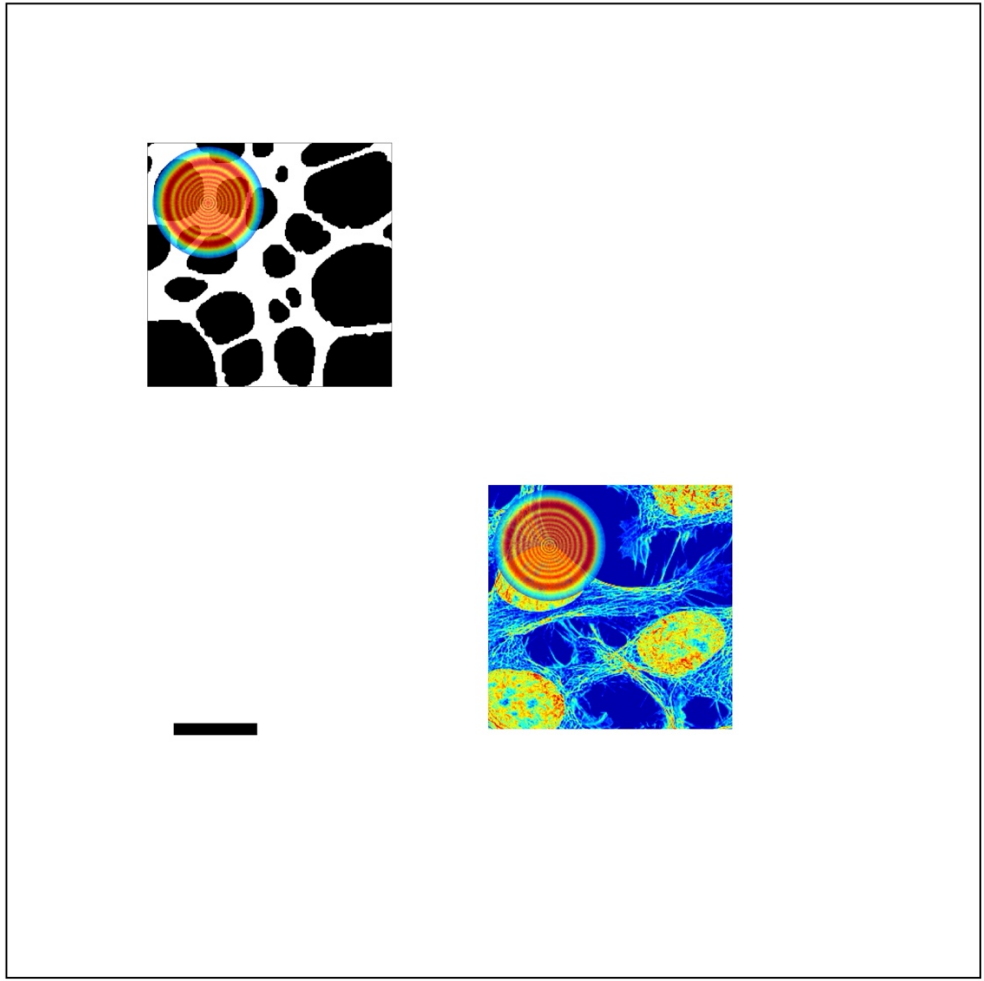

**Fig. S6**. Schematic of low-dose CDI used for the numerical experiment. The static structure consists of a 20-nm-thick Au pattern with a size of 2.92×2.92 μm$^2$. A simulated biological sample is HeLa cells with sharp features, a size of 2.92×2.92 μm$^2$ and a thickness of 300 nm. The black square indicates the whole reconstructed region. Scale bar, 1 μm.

# Supplemental Tables

**Table S1.** Parameters for the numerical experiments of in situ CDI.

| | |
|---|---|
| X-ray energy | 530 eV |
| Fluence on the static structure | 1.4e11 photons/$\mu m^2$ |
| Fluence on the dynamic structure | 1.75e3 – 3.5e8 photons/$\mu m^2$ |
| Size of the static structure | 2.92×2.92 $\mu m^2$ |
| Size of the dynamic structure | 2.92×2.92 $\mu m^2$ |
| Sample to detector distance | 5 cm |
| Pixel size of the detector | 10×10 $\mu m^2$ |
| # of pixels of the detector | 1024×1024 |
| Detector quantum efficiency | 100% |
| Oversampling ratio of static and dynamic structure | 1024×1024/(2×256×256) = 8 |
| Image pixel size | 11.4 nm |
| **Simulated sample parameters** | |
| A dynamic biological vesicle of protein complexes ($H_{50}C_{30}N_9O_{10}S_1$, density: 1.35 g/cm$^3$) immersed in water, where both the phase-shift and the absorption of sample is considered. | Thickness: 100 nm<br>Phase shift: 0.06 – 0.22 rad |
| | Thickness: 300 nm<br>Phase shift: 0.19 – 0.66 rad |
| | Thickness: 1 $\mu m$<br>Phase shift: 0.64 – 2.19 rad |

**Table S2.** Quantifying the spatial resolution of the phase-contrast perfect lens images and the in situ CDI and LoCDI reconstructions with FRC = 1/e.

| | | | | |
|---|---|---|---|---|
| Fig. 3 with a 300-nm-thick dynamic biological vesicle | | | | |
| X-ray fluence (photons/µm$^2$) | 3.5e5 | 3.5e6 | 3.5e7 | 3.5e8 |
| Perfect lens (nm) | 123.0 | 81.7 | 59.5 | 34.7 |
| In situ CDI (nm) | 100.1 | 70.2 | 49.0 | <22.8 |
| Fig. 4 with a 300-nm-thick biological sample of HeLa cells | | | | |
| X-ray Fluence (photons/µm$^2$) | 3.5e5 | 3.5e6 | 3.5e7 | 3.5e8 |
| Perfect lens (nm) | 75.5 | 25.7 | <22.8 | <22.8 |
| LoCDI (nm) | 34.5 | <22.8 | <22.8 | <22.8 |
| Ptychography (nm) | 180.7 | 93.6 | 44.1 | <22.8 |
| Fig. 5 LoCDI and ptychography experiment | | | | |
| Fluence (photons/mm$^2$) | 2.2e7 | 1.7e8 | 3.7e8 | 3.8e9 |
| LoCDI (µm) | 18.7 | 15.3 | 13.4 | 11.8 |
| Ptychography (µm) | 81.1 | 50.4 | 38.7 | 20.1 |
| Fig. S3 with a 100-nm-thick dynamic biological vesicle | | | | |
| X-ray Fluence (photons/µm$^2$) | 3.5e5 | 3.5e6 | 3.5e7 | 3.5e8 |
| Perfect lens (nm) | 182.8 | 101.6 | 70.2 | 50.5 |
| In situ CDI (nm) | 159.2 | 94.4 | 67.1 | 45.9 |
| Fig. S5 with a 1-µm-thick dynamic biological vesicle | | | | |
| X-ray Fluence (photons/µm$^2$) | 1.75e3 | 3.5e3 | 6.2e3 | 7.7e3 |
| In situ CDI (nm) | 189.9 | 155.6 | 131.2 | 123.3 |

**Table S3.** Parameters for the numerical experiments of LoCDI.

| | |
|---|---|
| X-ray energy | 530 eV |
| Size of the x-ray probe | 1.2 μm |
| Scan step size | 0.3 μm |
| # of the scan position | 10×10 |
| Size of a biological sample (HeLa cells) | 2.92×2.92 μm$^2$ |
| Thickness of the biological sample | 300 nm |
| Effective x-ray fluence on the biological sample | 3.5e5 – 3.5e8 photons/μm$^2$ |
| Size of a static structure (a 20-nm-thick Au pattern) | 2.92×2.92 μm$^2$ |
| Effective x-ray fluence on the static structure | 1.4e11 photons/μm$^2$ |
| Sample to detector distance | 5 cm |
| Pixel size of the detector | 10×10 μm$^2$ |
| # of pixels of the detector | 1024×1024 |
| Detector quantum efficiency | 100% |

**Video S1.** A dynamic biological vesicle model of protein complexes immersed in water, consisting of 20 frames, where circular features are due to the projection of the 3D membranes of the biological vesicle onto a 2D plane in real space.

**Video S2.** Phase-contrast images with perfect lenses and in situ CDI reconstructions of the dynamic biological vesicles with an x-ray fluence of 3.5e6 photons/μm$^2$. (a) The 100-nm-thick dynamic biological vesicle, consisting of 20 frames, obtained by phase-contrast microscopy with perfect lenses. (b) The 100-nm-thick dynamic biological vesicle reconstructed by in situ CDI. (c) The 300-nm-thick dynamic biological vesicle, consisting of 20 frames, obtained by phase-contrast microscopy with perfect lenses. (d) The 300-nm-thick dynamic biological vesicle reconstructed by in situ CDI.